\documentstyle[sprocl]{article}
\input{amssymb.sty}

\def\Journal#1#2#3#4{{#1} {\bf #2}, #3 (#4)}

\def\al{\alpha}

\def\be{\begin{equation}}
\def\ee{\end{equation}}
\def\bea{\begin{eqnarray}}
\def\eea{\end{eqnarray}}

\begin{document}

\title{
THE THEORY OF BINARY NONLINEARIZATION AND \\ITS APPLICATIONS TO SOLITON EQUATIONS}

\author{WEN-XIU MA}

\address{Department of Mathematics, City University of Hong Kong, 
Hong Kong, \\Kowloon, P. R. China
\\E-mail: mawx@cityu.edu.hk}


\maketitle\abstracts{
Binary symmetry constraints are applied to 
the nonlinearization of spectral problems and adjoint spectral problems
into so-called binary constrained flows, 
which provide candidates for 
finite-dimensional Liouville integrable Hamiltonian systems.
The resulting constraints on the potentials of spectral problems 
give rise to a kind of involutive solutions to zero curvature equations,
and thus the integrability by quadratures can be shown for zero curvature equations
once the corresponding binary constrained flows are proved to be integrable.
The whole process to carry out binary 
symmetry constraints is called binary nonlinearization. The principal task of 
binary nonlinearization is to expose the Liouville integrability for 
the resulting binary constrained flows, which can usually be achieved
as a consequence of the existence of 
hereditary recursion operators.  
The theory of binary nonlinearization is applied to 
the multi-wave interaction equations associated with a $4\times 4$ 
matrix spectral problem 
as an illustrative example.
The B\"acklund transformations resulted from symmetry constraints
are given for the multi-wave interaction equations, and thus a kind of involutive solutions 
is presented and the integrability by quadratures is shown
for the multi-wave interaction equations.}

\def \part {\partial}
\def \be {\begin{equation}}
\def \ee {\end{equation}}
\def \bea {\begin{eqnarray}}
\def \eea {\end{eqnarray}}
\def \ba {\begin{array}}
\def \ea {\end{array}}
\def \si {\sigma}
\def \al {\alpha}
\def \la {\lambda }
\def\D{\displaystyle}
\def\diag {\textrm{diag}}
\def \tr {\textrm{tr}}

\newcommand{\Z}{\mathbb{Z}}
\newcommand{\C}{\mathbb{C}}
\newcommand{\R}{{\mathbb{R}}}

\newtheorem{lemma}{Lemma}[section]
\newtheorem{theorem}{Theorem}[section]
\newtheorem{definition}{Definition}[section]
\newtheorem{proposition}{Proposition}[section]
\renewcommand{\theequation}{\thesection.\arabic{equation}}

\section{Introduction}
\setcounter{equation}{0}

The nonlinearization technique arose in the theory of soliton equations 
ten years ago \cite{Cao-SC1990}$^\textrm{-}\,$\cite{Ma-book1990}. 
One gradually realizes that 
it provides a powerful approach for analyzing 
soliton equations, in both continuous and discrete cases, especially for showing the integrability by quadratures for soliton equations.
The manipulation of nonlinearization
leads to finite-dimensional Liouville integrable systems 
(for example, see \cite{Ma-book1990}$^\textrm{-}\,$\cite{RagniscoW-JMP1994})
and also 
establish a bridge between
infinite-dimensional soliton equations and
finite-dimensional Liouville integrable systems 
\cite{AntonowiczW-JMP1992}$^\textrm{-}\,$\cite{CaoWG-JMP1999}.
What is more, 
it paves a method of separation of variables  
for soliton equations \cite{Sklyanin-PTPS1995}$^\textrm{-}\,$\cite{MaZ-krustal2000},
and thus the determination of finite-gap solutions to soliton equations
is transformed into solving the corresponding Jacobi inversions problems
\cite{ZengM-JMP1999,MaZ-krustal2000}.

Mathematically speaking, much excitement in the theory of nonlinearization
comes from a kind of specific symmetry constraints,
engendered from the variational derivative of the 
spectral parameter of spectral problems \cite{MaS-PLA1994,Ma-JPSJ1995},
which has a close relation with eigenfunctions of hereditary recursion operators
\cite{Ma-PA1995,MaFO-PA1996}.
Symmetry constraints are themselves a common conceptional umbrella under 
which one can manipulate both mono-nonlinearization 
\cite{Cao-SC1990} and binary nonlinearization \cite{MaS-PLA1994}, and 
the study of symmetry constraints is a crucial part of 
the kernel of the mathematical theory of nonlinearization.

Although 
it is not clear how to 
determine pairs of canonical variables to obtain 
Hamiltonian structures of constrained flows 
in a general case of mono-nonlinearization,
there exists a natural way for determining symplectic
structures to exhibit Hamiltonian forms for binary constrained flows
while doing binary nonlinearization \cite{MaF-book1996}.
Therefore, one can expect to establish
a rigorous theory for binary nonlinearization. 
In this paper, 
we would only like to discuss the procedure of manipulating 
binary nonlinearization. More considerations on the topics related to 
binary nonlinearization will be made elsewhere. 
  
The paper is structured as follows.
In section 2, a procedure for doing 
binary nonlinearization is depicted for general matrix
spectral problems.  
Then in section 3, 
an illustrative example
is carried out to make an application of the procedure to 
the multi-wave interaction equations associated with a $4\times 4$ matrix
spectral problem. Finally in section \ref{sec:summary_manila00},
a summary is given along with some concluding remarks.

\section{Binary Nonlinearization}
\setcounter{equation}{0}

In this section, we would like to 
describe the procedure of manipulating binary nonlinearization
for soliton equations
(see, for example, \cite{MaS-PLA1994,MaF-book1996}).
Let us start from  
a general matrix spectral problem
\begin{equation}
\phi_x=U\phi = U(u,\lambda)\phi,\ U=(U_{ij})_{r\times r},\ 
\phi=(\phi_1,\cdots,\phi_r)^T
\label{gsp}
\end{equation}
where $\la $ is a spectral parameter
 and $u=(u_1,\cdots,u_q)^T$ is a $q$-dimensional vector potential.
Associated with the spectral problem (\ref{gsp}),
suppose that for each integer $m\ge 0$, we have 
the evolution law for the eigenfunction $\phi$:
\be 
\phi_{t_m}=V^{(m)}\phi = V^{(m)}(u,u_x,\cdots;\lambda)\phi,\ 
V^{(m)}=(V^{(m)}_{ij})_{r\times r }. \label{gassosp}
\ee  
If the Gateaux derivative $U'$ of $U$ with the potential $u$
is injective, then  
the isospectral ($\la _{t_m}=0$) zero curvature equation
\be U_{t_m}-V_x^{(m)}+[U,V^{(m)}]=0\label{eq:zce}\ee 
will determine \cite{Ma-JPA1993}
an evolution equation 
\be 
u_{t_m}=X_m(u)=JG_m=J\frac{\delta {\tilde H_m}}{\delta u},\ u=u(x,t_m),
\label{eq:eeofmth}
\ee 
where $X_m$ depends on the potential $u$
and its spatial derivatives of $u$
up to some finite order, 
$J(u)$ is a Hamiltonian operator, and 
${\tilde H}_m(u)$ is  
a Hamiltonian functional. 
Evidently, the compatability condition of 
the adjoint spectral problem
and the adjoint associated spectral problem 
\be 
\psi _x=-U^T(u,\la )\psi ,\ 
\psi _{t_m} =-V^{(m)T}(u,\la )\psi,\ 
\psi =(\psi _1,\cdots ,\psi_r)^T
\ee 
is still the same as the zero curvature equation (\ref{eq:zce}), guaranteed by 
the evolution equation $u_{t_m}=X_m(u)$ defined by (\ref{eq:eeofmth}).

It has been pointed out \cite{MaS-PLA1994,Ma-JPSJ1995} that 
\[
J\D\frac {\delta \lambda }{\delta u}=E^{-1} J\psi ^T
\frac{\part U(u,\la )}{\part u}\phi,\ 
E=-{\int_\Omega \psi^T\frac{\partial U(u,\la )}{\partial \la }\phi\, dx,}
\]
is a  
symmetry of the evolution equation (\ref{eq:eeofmth}),
where $E$ is called the normalized constant, and $\Omega =(0,T)$ if $u$ is assumed to 
be periodic with period $T$ and $\Omega =(-\infty ,\infty )$ if 
$u$ is assumed to 
belong to the Schwartz space. 
Let us now introduce
$N$ distinct eigenvalues $\la _1,\cdots,\la _N$,
and so we have
\be 
\phi^{(s)}_x=U(u,\la _s)\phi^{(s)}, \ \psi^{(s)}_x=-U^T(u,\la _s)\psi^{(s)},
\ 1\le s\le N;\qquad 
\label{gxpartofcf}
\ee
\be   
\phi^{(s)}_{t_m}=V^{(m)}(u,\la _s)\phi^{(s)}, \ \psi^{(s)}_{t_m}
=-V^{(m)T}(u,\la _s)\psi^{(s)},\ 1\le s\le N;
\label{gtpartofcf}
\ee 
where the corresponding eigenfunctions and adjoint eigenfunctions 
are denoted by $\phi^{(s)}$ and $\psi^{(s)}$, $1\le s\le N$.
Suppose that the covariant
$G_{m_0}$ is Lie point, and then
the so-called binary Bargmann symmetry constraint reads as
\be
X_{m_0}=\sum_{s=1}^N E_s\mu _s
J\frac {\delta \la _s}{\delta u},\ \textrm{i.e.,} \ 
JG_{m_0}=
J\sum_{s=1}^N\mu_s \psi^{(s)T}\frac {\part U(u,\la _s)}{\part u}\phi^{(s)},
\label{gsy} \ee 
where 
$\mu _s,\ 1\le s\le N,$ are arbitrary non-zero constants,
and $E_s,\ 1\le s\le N,$ are $N$ normalized constants.
The right-hand side of the binary symmetry constraint (\ref{gsy}) is 
a linear combination of $N$ symmetries 
$
E_sJ\D{\delta \lambda_s }/{\delta u},\ 1\le s\le N$,
and 
the requirement of $\mu _s\ne 0,\ 1\le s\le N$, is natural,
since 
the symmetry $J\D{\delta \lambda_s }/{\delta u}$ is just not involved in our symmetry 
constraint (\ref{gsy}) if some constant $\mu_s=0$.
It is also worthy to mention that 
$\phi^{(s)}$ and $\psi^{(s)}$ can not be expressed 
in terms of $x$, $u$ and spatial derivatives of $u$ to some finite order,
and thus such symmetries (or the corresponding covariants
$\D{\delta \lambda_s }/{\delta u},\ 1\le s\le N$)
are not Lie point, contact or Lie-B\"acklund
symmetries. 

Let us now assume that we can solve 
the symmetry constraint (\ref{gsy}) for $u$:
\be u=\widetilde u(\phi^{(1)},\cdots,\phi^{(N)};\psi^{(1)},\cdots,\psi^{(N)}).
\ee 
Replacing $u$ with $\widetilde u$ in 
the system ({\ref{gxpartofcf}) and 
the system (\ref{gtpartofcf}), we obtain the so-called spatial binary constrained flow:
\be 
\phi^{(s)}_x=U(\widetilde u,\la _s)\phi^{(s)}, \ \psi^{(s)}_x=-U^T(\widetilde u,\la _s)\psi^{(s)},
\ 1\le s\le N,\qquad 
\label{xpartofcf}
\ee 
and 
the so-called temporal
binary constrained flow
\be 
\phi^{(s)}_{t_m}=V^{(m)}(\widetilde u,\la _s)\phi^{(s)}, \ \psi^{(s)}_{t_m}
=-V^{(m)T}(\widetilde u,\la _s)\psi^{(s)}, \ 1\le s\le N.
\label{tpartofcf}
\ee 
Note that the spatial and temporal binary constrained flows 
(\ref{xpartofcf}) and (\ref{tpartofcf})
are nonlinear with respect to 
eigenfunctions and adjoint eigenfunctions,
but the original spectral problems ({\ref{gxpartofcf}) and ({\ref{gtpartofcf})
are linear.
Moreover, the spatial binary constrained flow (\ref{xpartofcf})
 is a system of ordinary differential equations,
and
the temporal binary constrained flow (\ref{tpartofcf})
 is usually a system of 
partial differential equations but it can be transformed
into a system of ordinary differential equations under the control of (\ref{xpartofcf}).

The principal task in the theory of binary nonlinearization is to show that 
the spatial binary constrained flow (\ref{xpartofcf})
and the temporal binary constrained flow (\ref{tpartofcf}) 
under the control of (\ref{xpartofcf})
are Liouville integrable, and thus the integrability by quadratures can be shown for evolution equations possessing zero curvature representations.  
The main tools to do this are
Lax representations generated from the stationary zero curvature equations 
and $r$-matrix formulations of Lax operators \cite{Ma-WCNA2000}.
These will form 
the important part of the theory of binary nonlinearization and 
will be shed right on in our example.

Now if $\phi^{(s)}$ and $\psi ^{(s)}$, $1\le s\le N$,
solve two binary constrained flows (\ref{xpartofcf}) and (\ref{tpartofcf}) simultaneously,  
then $u=\widetilde u$
will present a solution to 
the evolution equation (\ref{eq:eeofmth}),
since the evolution equation (\ref{eq:eeofmth}) with the potential $u=\widetilde u$
is the compatability of 
two binary constrained flows (\ref{xpartofcf}) and (\ref{tpartofcf}). 
It also follows that the evolution equation 
$u_{t_m}=X_m(u)$ is decomposed into two finite-dimensional Liouville integrable 
systems, and   
the constraint $u=\widetilde u$ gives rise to 
a B\"acklund transformation between the evolution 
equation and 
the resulting finite-dimensional Liouville integrable 
systems \cite{MaG-book1999}.
Furthermore, two binary constrained flows (\ref{xpartofcf}) and (\ref{tpartofcf})
can be solved by separation of varaibles and all what we need to do 
upon obtaining separated variables is to solve the Jacobi inversion problems
\cite{MaZ-krustal2000}.
This whole process to carry out symmetry constraints 
is called binary nonlinearization \cite{Ma-PA1995,MaF-book1996}. 

\section{Applications to Soliton Equations}
\setcounter{equation}{0}


By a soliton hierarchy we means a hierarchy of evolution equations 
with a recursion relation
as follows
\be 
u_{t_m}=X_m(u)=\Phi ^mX_0=JG_m=J\frac{\delta {\tilde H_m}}{\delta u},\ m\ge 0,
\label{sh}
\ee 
where $u$ is the vector potential $ u=u(x,t_m)$, 
$X_m$, $m\ge 0$, depend on the potentional 
$u$ and its spatial derivatives of $u$
up to some finite order, 
$\Phi $ is a herediatry common recursion operator, 
$J(u)$ is a Hamiltonian operator, and 
${\tilde H}_m(u)$, $m\ge 0$, are  
the Hamiltonian functionals. 
Moreover, we have 
\bea && 
[X_m,X_n]:=\frac {\partial }{\partial \varepsilon}
\Bigl(X_m(u+\varepsilon X_n
- X_n(u+\varepsilon X_m )\Bigl)
\Bigl.\Bigr|_{\varepsilon=0}=0,\ m,n\ge 0,\qquad \\ &&
\{\tilde {H}_m,\tilde {H}_n\}_J:=\int ( \frac{\delta \tilde {H}_m}{\delta u})^TJ
\frac{\delta \tilde {H}_n}{\delta u}\, dx= 0,\ m,n\ge 0,
\eea 
where $[\cdot,\cdot]$ and $\{\cdot,\cdot\}_J$ are called 
the commutator of vector fields and the Poisson bracket assocaited with the Hamiltonian 
 operator $J$, resprectively.
It follows that each equation $u_{t_m}=X_m$ in the soliton hierarhy (\ref{sh})
has infinitely many symmetries $\{X_n\}_{n=0}^\infty$ and infinitely many conserved functionals $\{\tilde {H}_n\}_{n=0}^\infty$.
This property can be obtained from a bi-Hamiltonian formaultion,
and its natrural bi-Hamiltonian formaulation for the hierarchy (\ref{sh})
is formed by $J$ and $M=J\Psi$ 
if they are a Hamiltonian pair, where 
$\Psi$ denotes the adjoint operator of $\Phi$, i.e., 
$\Psi=\Phi^\dagger$. The above property can aslo be obtained from 
a weaker condition that $M$ is a symmetric operator ($M^\dagger =M$), 
i.e,
$ \Phi J = J\Psi $ \cite{Tu-JMP1989}. 

A soliton hierarchy can usually be constructed from 
a hierarchy of zero curvature equations as the compatability 
conditions of
a given matrix spectral problem (\ref{gsp}) and 
its associated spectral porblems (\ref{gassosp}) 
\cite{NovikovMPZ-book1984,AblowitzC-book1991}. 
The hereditary recursion operator can be generated from 
the matrix spectral problem (\ref{gsp}),
and the Hamilotnian formulations can be determined by a trace identity
\cite{Tu-JMP1989,Tu-JPA1989}. All this information shows
that we can have everything for manipulating binary nonlinearization
for our soliton hierarchy (\ref{sh}). 
Applications of binary nonlinearization have been made 
to the KdV, the MKdV, the AKNS, the Kaup-Newell, the Dirac, the WKI soliton hierarchies 
and so on \cite{MaS-PLA1994}\hspace{-0.45em}
$^\textrm{,}\,$\cite{Ma-JPSJ1995}$^\textrm{-}\,$\cite{MaF-book1996}
 \cite{Xu-CPL1995}$^\textrm{-}\,$\cite{LiM-CSF2000}.
In what follows,   
an illustrative example of the multi-wave interaction equations associate with a $4\times
4$ matrix spectral porblem 
will be carried out for making applications 
of the procedure of binary nonlinearization.

\subsection{Multi-wave Interaction Equations}

Let us start from a 4$\times $4 matrix spectral problem: 
\be \phi _x=U\left( u,\lambda \right) \phi ,\ U=\left( 
\begin{array}{cccc}
\al _1\lambda  & u_{12} & u_{13} &u_{14} \\ 
u_{21} & \al _2 \la & u_{23} &u_{24} \\ 
u_{31} & u_{32} & \al _3\la &u_{34}\\
u_{41} & u_{42} & u_{43}& \al _4
\la   
\end{array}
\right)=\lambda U_0  +U_1 ,\ \phi =\left( 
\begin{array}{c}
\phi _1 \\ 
\phi _2 \\ 
\phi _3\\
\phi_4 
\end{array}
\right), 
\label{eq:4spNWIEs} \ \ee 
where $U_0=\textrm{diag}(\al _1,\al _2,\al _3,\al_4)$,
 and $\al _1,\al _2,\al _3,\al _4$ 
are distinct constants, and the potential $u$ 
is defined by 
\be u=\rho (U_1)=(u_{21},u_{12},u_{31},u_{13},u_{41},u_{32},u_{23},
u_{14},u_{42},u_{24},u_{43},u_{34})^T .
\label{eq:defofrho}\ee 
The associated spectral problem is chosen as 
\be \phi _t= V\left( u,\lambda \right) \phi ,\ V=\left( 
\begin{array}{cccc}
\beta _1\lambda  & v_{12} & v_{13}  &v_{14}\\ 
v_{21} & \beta _2 \la & v_{23}  &v_{24}\\ 
v_{31} & v_{32} & \beta _3 \la &v_{34}\\
  v_{41} & v_{42} & v_{43}& \beta _4 \la 
\end{array}
\right)=\lambda V_0 +V_1 ,
\label{eq:4aspNWIEs} \ee 
where $V_0=\textrm{diag}(\beta _1,\beta _2,\beta _3,\beta _4)$, and $\beta _1,\beta _2,\beta _3,
\beta_4$ 
are distinct constants.
Then the isospectral ($\la _t=0$) compatibility condition 
$ U_{t}-V_x+[U,V]=0$ 
of the spectral problem
(\ref{eq:4spNWIEs}) and the associated spectral problem (\ref{eq:4aspNWIEs}) 
gives rise to  
\be U_{1t}-V_{1x}+[U_1,V_1]=0,\ [U_0,V_1]=[V_0,U_1].
\label{eq:matrixformof4MWIEs}
\ee 
These two matrix equations lead to the following multi-wave interaction equations 
\be 
u_{ij,t}=\frac{\beta_i-\beta_j}{\alpha_i-\alpha_j}u_{ij,x}
+\sum^4_{\stackrel{k=1}{k\not=i,j}}(\frac{\beta_k-\beta_i}{\alpha_k-\alpha_i}
-\frac{\beta_k-\beta_j}{\alpha_k-\alpha_j})u_{ik}u_{kj},\  1\leq i\ne j\leq 4, \ \,
\label{eq:4MWIEs}\ee 
which contain 
a couple of 
physically important nonlinear models as special reductions \cite{AblowitzC-book1991}.
Note that the compatability condition of
the adjoint spectral problem
and adjoint associated spectral problem: 
\be \psi_x=-U^T(u,\la )\psi ,\ \psi _t=-V^T(u,\la )\psi,\ \psi=(\psi _1,\psi_2,\psi_3,\psi_4)^T \ee
still gives rise to the above multi-wave interaction equations.
It is easy to find that the multi-wave interaction equations 
(\ref{eq:4MWIEs}) have the Hamiltonian structure
\be 
u_{ij,t}=J\frac {\delta {\tilde H}}{\delta u_{ij}} ,\ 1\le i\ne j\le 4, \label{eq:Hfof4MWIEs}\ee 
where the Hamiltonian operator $J$ 
reads as
\bea && 
J=\textrm{diag}\Bigl( (\alpha_1-\alpha_2)\sigma_0, (\alpha_1-\alpha_3)\sigma_0,
(\alpha_1-\alpha_4)\sigma_0, \nonumber \\ \quad  \qquad &&
(\alpha_2-\alpha_3)\sigma_0,(\alpha_2-\alpha_4)\sigma_0,(\alpha_3-\alpha_4)\sigma_0\Bigl),\ 
\sigma_0=\left( \begin{array} {cc} 0&\, 1 \vspace{2mm} \\ -1&\, 0 \end{array}
\right ). \qquad 
\eea 
The Hamiltonian function $\tilde {H}$ in (\ref{eq:Hfof4MWIEs}) 
can directly be computed but it is omitted
here due to its complicated form.
We remark that the soliton hierarchy containing higher-order 
multi-wave interaction equations 
can also be generated from the spectral problem 
(\ref{eq:4aspNWIEs}), but in this report we focus on the multi-wave interaction equations 
(\ref{eq:4MWIEs}).

\subsection{Binary Constrained Flows}

Let us now construct  
binary constrained flows of the multi-wave interaction equations (\ref{eq:4MWIEs}).
Based on the general procedure of binary nonlinearization, 
the symmetry of (\ref{eq:4MWIEs}) generated from the conserved functional $\lambda(u) $ 
reads as
\be EJ\frac {\delta \la }{\delta u}=E 
\rho([U_0,\rho^{-1}(\frac {\delta \lambda }{\delta u})])
=\rho([U_0,
 \rho^{-1}(\psi ^T\frac {\partial U(u,\lambda )}{\partial u}\phi )]),
\ee 
where $E$ is the normalized constant and $\rho $ is the mapping defined by (\ref{eq:defofrho}).
Upon introducing
$N$ distinct eigenvalues $\la _1,\cdots, \la _N$,
we obtain $N$ replicas of Lax systems:
\be  \phi^{(s)} _{x}=U(u,\la _s)
\phi ^{(s)},\  
\psi ^{(s)}_{x}=-U^T(u,\la _s)
\psi^{(s)},\ 1\le s\le N;
\label{eq:originalxpartofbinaryLaxsystems}
 \ee \be 
\phi^{(s)} _{t}=V(u,\la _s)
\phi ^{(s)},\  
\psi ^{(s)}_{t}=-V^T(u,\la _s)
\psi^{(s)},\ 1\le s\le N; \label{eq:originaltpartofbinaryLaxsystems}
\ee  
where the eigenfunctions 
$\phi^{(s)}$ and the adjoint eigenfunctions
$\psi^{(s)}$ are assumed to be denoted by 
\be 
\phi^{(s)}=(\phi_{1s},\phi_{2s},\phi_{3s},\phi_{4s})^T,
\ \psi^{(s)}=(\psi_{1s},\psi_{2s},\psi_{3s},\psi_{4s})^T,\ 1\le s\le N.\ee 
These $N$ eigenvalues
lead to a more general symmetry of (\ref{eq:4MWIEs})
\[  Z_0:=
\rho ([U_0,\rho^{-1}(\D \sum_{s=1}^N \mu_s \psi ^{(s)T} \frac{\partial U(u,\lambda _s)}{\partial u}\phi^{(s)})])= \rho([U_0,\D \sum_{s=1}^N \mu_s \phi ^{(s)}\psi^{(s)T}]), \]
where the $\mu_s$'s are arbitrary non-zero constants. 
On the other hand, we can find by inspection that 
the multi-wave interaction equations 
(\ref{eq:4MWIEs}) have a Lie point symmetry  
\[ Y_0:= \rho ([\Gamma, U_1]),\ 
\Gamma=\diag (\gamma_1,\gamma_2,\gamma _3,\gamma _4),\ 1\le i\ne j\le 4, \]
where $\gamma_1,\gamma_2,\gamma_3,\gamma _4$ are distinct constants.
In fact, we can directly prove that both $Y_0$ and $Z_0$ are two symmetries of 
the multi-wave interaction equations 
(\ref{eq:4MWIEs}). 
Define
\[ \delta U_1=[\Gamma, U_1]\ \textrm{or}\  [U_0,\sum_{s=1}^N \mu_s 
\phi ^{(s)}\psi^{(s)T}],\]
then $[U_0,\delta V_1]=[V_0,\delta U_1]$ uniquely determines
\[ \delta V_1=[\Gamma, V_1]\ \textrm{or}\ [V_0,\sum_{s=1}^N\mu_s 
\phi ^{(s)}\psi^{(s)T}].\]  
Now it follows from (\ref{eq:matrixformof4MWIEs}) that
the symmetry problem only requires to verify that 
\[ (\delta U_1,\delta V_1)=([ \Gamma, U_1],[\Gamma, V_1])\ \textrm{or} \ 
([U_0,\D \sum_{s=1}^N\mu_s 
\phi ^{(s)}\psi^{(s)T}],[V_0,\sum_{s=1}^N\mu_s 
\phi ^{(s)}\psi^{(s)T}])\ \] 
satisfies the linearized system of the multi-wave interaction equations 
(\ref{eq:4MWIEs}): 
\be 
 (\delta U_1)_t-(\delta V_1)_x+[\delta U_1,V_1]+[U_1,\delta V_1]=0,\ee
which can easily be proved. 

Therefore, a binary symmetry constraint of (\ref{eq:4MWIEs}) 
can be taken as 
\be Y_0=Z_0,\ \textrm{i.e.,}\ 
[\Gamma, U_1]=
[U_0,\sum_{s=1}^N\mu_s
\phi ^{(s)}\psi^{(s)T} ]. \label{eq:bscof4MWIEs}\ee 
Solving this equation for $u$, we obtain the required constraints on the potentials
\be  u_{ij}=\tilde u_{ij}:=
\frac{\alpha_i-\alpha_j}{\gamma_i-\gamma_j}\langle\Phi_i,B\Psi_j\rangle,\ 
1\leq i\ne j\leq 4, \label{eq:defoftildeu}\ee 
where $\langle\cdot,\cdot\rangle$ denotes the standard inner-product 
of ${\R}^N$, $B$ is defined by 
\be B=\textrm{diag}(\mu_1,\mu_2,\cdots,\mu_N),\label{eq:defofB}\ee
and $\Phi_i$, $\Psi_i$ are defined by
\be \Phi_{i}=(\phi_{i1},\cdots,\phi_{iN})^T,\ 
\Psi_{i}=(\psi_{i1},\cdots,\psi_{iN})^T,\ 1\le i\le 4.
\label{eq:defofPhi_iandPsi_i}
\ee
The substitution of $u$ with 
\[
 {\tilde u}=({\tilde u}_{21},{\tilde u}_{12},
{\tilde u}_{31},{\tilde u}_{13},{\tilde u}_{41},{\tilde u}_{32},{\tilde u}_{23},
{\tilde u}_{14},{\tilde u}_{42},{\tilde u}_{24},{\tilde u}_{43},{\tilde u}_{34} )\]
into the binary Lax systems (\ref{eq:originalxpartofbinaryLaxsystems})
and (\ref{eq:originaltpartofbinaryLaxsystems})
yields the so-called binary constrained flows:
\be  
\phi^{(s)} _{x}=U(\tilde u ,\la _s) \phi ^{(s)},\ 
\psi^{(s)}_{x}=-U^T(\tilde u ,\la _s) \psi^{(s)},\ 1\le s\le N;\label{eq:xcfof4NWIEs}
\ee \be  \phi ^{(s)}_{t}=V(\tilde u ,\la _s) \phi ^{(s)} 
,\ \psi^{(s)}_{t}=-V^T(\tilde u ,\la _s) \psi ^{(s)},\ 1\le s\le N;\label{eq:tcfof4NWIEs}\ee   
which are two systems of ordinary differential equations, since $V$ doesn't involve 
any spatial derivative of $u$.

\subsection{Liouville Integrability}

In order to analyze the Liouville integrability of the above two binary constrained flows,
let us first introduce
a symplectic structure over ${\mathbb{R}} ^{8N}$:
\be 
\omega ^2= \sum_{i=1}^4 B d\Phi _i\wedge d\Psi _i=
\sum_{i=1}^4\sum_{s=1}^N \mu _s  d\phi_{is}\wedge d\psi _{is},
\label{symplecticform}
\ee 
where $B$ is defined by
(\ref{eq:defofB}) and $\Phi_i$ and $\Psi_i$ are defined by 
(\ref{eq:defofPhi_iandPsi_i}).
Then the associated Poisson bracket reads as
\bea && \{f,g\}=\omega^2 (Idg,Idf)=\sum_{i=1}^4(\langle \frac {\part f}{\part \Psi_i},
B^{-1}\frac {\part g}{\part \Phi_i} \rangle -
\langle \frac {\part f}{\part \Phi_i},
B^{-1}\frac {\part g}{\part \Psi_i} \rangle  )\nonumber \\ &&
=\sum_{i=1}^4\sum_{s=1}^N
\mu _s^{-1} \Bigl(\frac{\part f}{\part \psi_{is}} \frac{\part g}{\part \phi_{is}}
-\frac{\part f}{\part \phi_{is}} \frac{\part g}{\part \psi_{is}}
\Bigr),\ f,g\in C^\infty ({\mathbb{R}}^{ 8 N }), \label{eq:pbof4MWIEs}\eea  
where the vector field $Idf$ is uniquely determined by 
\[\omega^2(X,Idf)=df(X),\ X\in T({\mathbb{R}}^{ 8 N}).  \]
A Hamiltonian system with a Hamiltonian $H$ defined over 
this symplectic manifold $({\mathbb{R}}^{ 8 N},\omega ^{2})$ is given by 
\be 
\Phi_{it}=\{\Phi_{i},H\}=-B^{-1}\frac {\part H}{\part \Psi _i},\ \Psi_{it}=
\{\Psi_{i},H\}=
B^{-1}\frac {\part H}{\part \Phi_i},\ 1\le i\le 4,
\label{eq:gHamiltonianform}
\ee 
where $t$ is assumed to be the evolution variable.

In order to prove that two binary constrained flows 
are Liouville integrable, we need generate 
their Hamiltonian structures and integrals of motion. 
A direct computation can verify the following theorem.
\begin{theorem}
The spatial binary constrained flow (\ref{eq:xcfof4NWIEs})
and the temporal binary constrained flow (\ref{eq:tcfof4NWIEs})
are Hamiltonian systems with the evolution variables $x$, $t$ and 
 the Hamiltonian functions 
\bea  
H^x&=&- \sum^4_{k=1}\alpha_k\langle A    \Phi_k,B\Psi_k\rangle
-\sum_{1\leq k<l\leq 4}\frac{\alpha_k-\alpha_l}{\beta_k-\beta_l}\langle\Phi_k,B\Psi_l\rangle\langle\Phi_l,B\Psi_k\rangle,
\label{eq:defofH^x}
\\
H^t&=&- \sum_{k=1}^4\beta _k\langle A   \Phi _k,B\Psi_k\rangle
-\sum_{1\le k<l\le 4}\frac{\beta _k-\beta_l}{\gamma _k-\gamma _l}
\langle\Phi _k,B\Psi _l\rangle\langle\Phi _l,B\Psi_k\rangle, 
\label{eq:defofH^t}\qquad 
\eea  
respectively, where the matrix $A$ is given by 
\be
A=\textrm{diag}(\lambda _1,\lambda _2,\cdots ,\lambda _N).\ee
\end{theorem}

It is well known that 
Lax representations can engender integrals of motion.
For the binary constrained flows 
(\ref{eq:xcfof4NWIEs})
and (\ref{eq:tcfof4NWIEs}),
let us introduce a Lax operator $L(\la )$ as follows 
\bea && 
L(\la ) =(L_{ij}(\la ))_{4\times 4}= \Gamma+D(\la )=
\textrm{diag}(\gamma _1,\gamma_2,\gamma_3,\gamma_4)
+D(\la ),
 \label{eq:1ofdefofL(lambda)of4NWIEs}\\
&& D(\la )=(D_{ij}(\la ))_{4\times 4},\ D_{ij}(\la )= 
\D \sum_{l=1}^{N}
\frac{\mu_l \phi_{il}\psi_{jl}}{\lambda -\lambda _l},\ 1\le i,j\le 4.
\label{eq:2ofdefofL(lambda)of4NWIEs} \eea 
Now if we define 
\be \widetilde {U}(\la )=U(\tilde u,\la ),\ 
 \widetilde {V}
(\la )= V(\tilde u,\la ),\ee 
then we obtain the following result.
\begin{theorem}
The binary constrained flows 
(\ref{eq:xcfof4NWIEs})
and (\ref{eq:tcfof4NWIEs}) have 
the following Lax representations:
\be 
(L(\la ))_x=[\widetilde{U}(\la ),L(\la )],\ 
(L(\la ))_{t}=[
\widetilde{V}(\la ),L(\la )],
\label{eq:Laxrepresentationsofcfsof4NWIEs}
\ee 
respectively.
\end{theorem}



An $r$-matrix formulation can also be directly shown for the Lax operator 
$L(\la )$ defined by 
(\ref{eq:1ofdefofL(lambda)of4NWIEs}) and
(\ref{eq:2ofdefofL(lambda)of4NWIEs}).

\begin{theorem} The Lax operator $L(\lambda)$ defined by 
(\ref{eq:1ofdefofL(lambda)of4NWIEs}) and
(\ref{eq:2ofdefofL(lambda)of4NWIEs})
has the following $r$-matrix formulation
\be 
\{L(\lambda)\stackrel{\otimes}{,}
L(\mu)\}=\left[\frac{1}{\mu-\lambda}{\mathcal P},
L_{1}(\lambda)+L_{2}(\mu)\right] , \ 
 {\mathcal P}= \sum_{i,j=1}^4E_{ij}\otimes E_{ji},
\label{eq:rformulationofLaxoperatorof4NWIEs}\ee
where 
$L_1(\la )=L(\la )\otimes I_4,\ L_2(\mu )=I_4\otimes L(\mu ), 
$  
and 
\[
(E_{ij})_{kl}=\delta_{ik}\delta_{jl},\ 
\{L(\lambda)\stackrel{\otimes}{,}
L(\mu) \}_{ij,kl}=\{L_{ik}(\la ),L_{jl}(\mu )\},\ 1\le i,j,k,l\le 4.
\]
\end{theorem}

Now first from the Lax representations in
(\ref{eq:Laxrepresentationsofcfsof4NWIEs}), we obtain
\[ 
(\nu I_4-L(\la ))_x=[ \widetilde{U}(\la ),\nu I_4-L(\la )],\ 
(\nu I_4-L(\la ))_t=[ \widetilde{V}(\la ),\nu I_4-L(\la )],\ \]  
where $\nu $ is a parameter,
and thus \cite{Tu-JPA1989}
\[ (\det (\nu I_4-L(\la )))_x=0,\ (\det (\nu I_4-L(\la )))_t=0.\]
Second from the $r$-matrix formulation (\ref{eq:rformulationofLaxoperatorof4NWIEs}), we 
can obtain  
\[ \{\tr L^k(\lambda ),\tr L^l(\mu )\}=0,\  k,l\ge 1. \]
Expand the determinant of the matrix $\nu I_4-L(\la )$ as
\be \det(\nu I_4-L(\lambda))=\nu^4-{\mathcal F}_1(\lambda)\nu^3
+{\mathcal F}_2(\lambda)\nu^2
-{\mathcal F}_3(\lambda)\nu + {\mathcal F}_4(\lambda),
\ee 
where by Newton's identities on 
elementary symmetric polynomials, we have 
\[  \ba {rl} {\mathcal F}_1(\lambda)=&\tr L(\lambda),\ {\mathcal F}_2(\lambda)
=\frac 1 2((\tr L(\lambda))^2-
\tr L^2(\lambda )),
\vspace{2mm}\\
{\mathcal F}_3(\lambda)=&\frac 1 6 (\tr L(\lambda))^3+\frac 1 3\tr L^3(\lambda)
-\frac 1 2(\tr L(\lambda))\tr L^2(\lambda),
\vspace{2mm}\\
{\mathcal F}_4(\lambda)=&\det L(\lambda )=\frac 1 {24} (\tr L(\lambda))^4-
\frac 1 4 \tr L^4(\lambda) +\frac 18 (\tr L^2(\lambda))^2
\vspace{2mm}\\ &
-\frac 1 4(\tr L(\lambda))^2\tr L^2(\lambda)+\frac 13 (\tr L(\lambda))\tr L^3(\lambda)
.\ea \] 
Therefore, it follows that 
\be
({\mathcal F}_i(\la ))_x=0,\ ({\mathcal F}_i(\la ))_t=0,
\  \{{\mathcal F}_i(\la ),{\mathcal F}_j(\mu )\}=0,\ 1\le i,j\le 4.  
\label{eq:involutivepropertyofintegralsofmotion}\ee  
Make a further expansion of ${\mathcal F}_i(\la )$ as follows
\be  {\mathcal F}_i(\la )=\sum_{k\ge 0}F_{ik}\la ^{-k},\ 1\le i\le 4,\ee 
where obviously the $F_{i0}$'s are all constants.
Then it follows from (\ref{eq:involutivepropertyofintegralsofmotion})
that   
\be 
(F_{ik})_x=0,\ (F_{ik})_t=0,\ 
\{F_{ik},F_{jl}\}=0,\ 1\le i,j\le 4,\ k,l\ge 0,\label{eq:ipofF_{il}F_{jk}}\ee 
Therefore, 
two binary constrained flows (\ref{eq:xcfof4NWIEs})
and (\ref{eq:tcfof4NWIEs}) have the common integrals of 
motion: $F_{il},\ 1\le i\le 4,\ l\ge 1,$ which are in involution in pairs under the Poisson
bracket (\ref{eq:pbof4MWIEs}).
Now it is direct to 
show the following theorem on the Liouville integrability of two binary constrained flows
(\ref{eq:xcfof4NWIEs})
and (\ref{eq:tcfof4NWIEs}).

\begin{theorem} 
Two binary constrained flows (\ref{eq:xcfof4NWIEs})
and (\ref{eq:tcfof4NWIEs})
have the common integrals of motion:
$F_{il},\ 1\le i\le 4,\ l\ge 1$,
which are in involution in pairs under the Poisson
bracket (\ref{eq:pbof4MWIEs}), and 
of which the functions $F_{is},\ 1\le i\le 4,\ 1\le s\le N,$
are functionally independent
over a dense open subset of $\R ^{8N}$. Thus, 
two binary constrained flows (\ref{eq:xcfof4NWIEs})
and (\ref{eq:tcfof4NWIEs}) are Liouville integrable.
\end{theorem}

\subsection{Involutive Solutions}

Note that 
 under the constraints on the potentials (\ref{eq:defoftildeu}),
the compatability condition of 
(\ref{eq:originalxpartofbinaryLaxsystems}) and 
(\ref{eq:originaltpartofbinaryLaxsystems})
is still only the multi-wave interaction equations (\ref{eq:4MWIEs}).
Thus, solutions $(\Phi_i(x,t),\Psi_i(x,t))$ to two binary constrained flows (\ref{eq:xcfof4NWIEs}) and (\ref{eq:tcfof4NWIEs})
will present solutions to the multi-wave interaction equations (\ref{eq:4MWIEs}),
namely,
\be 
u_{ij}(x,t)=\frac {\al _i-\al _j}{\gamma _i -\gamma  _j}
\langle\Phi_i(x,t) ,
B\Psi_j(x,t)\rangle,
\ 1\le i\ne j\le 4,\label{eq:involutivesolutionsof4MWIEs}\ee 
which also shows the integrability by quadratures for the 
multi-wave interaction equations (\ref{eq:4MWIEs})
because $\Phi_i$ and $\Psi_i$, $1\le i\le 4$, can be determined 
by quadratures.

It is also easy to observe that 
under the Poisson bracket (\ref{eq:pbof4MWIEs}), two Hamiltonians $H^x$ and $H^t$ 
defined by (\ref{eq:defofH^t}) and (\ref{eq:defofH^t}) 
commute, namely  
\be \{H^x,H^t\}= \sum_{i=1}^4 \Bigl(\langle \frac {\part H^x}{\part \Psi_i}
,B^{-1}\frac {\part H^t}{\part \Phi_i}\rangle - \langle
\frac {\part H^x}{\part \Phi_i},
B^{-1}\frac {\part H^t}{\part \Psi_i}
\rangle \Bigr) = 0.\ee 
It follows that the above solutions determined by (\ref{eq:involutivesolutionsof4MWIEs})
give rise to a kind of involutive solutions to the multi-wave interaction equations
(\ref{eq:4MWIEs}). Let us denote two Hamiltonian flows of
(\ref{eq:xcfof4NWIEs}) and (\ref{eq:tcfof4NWIEs}) 
by $g_x^{H^x }$ and $g_{t}^{H^t}$ respectively.
Then the resulting involutive solutions can be written as
\[ \ba {rcl}    
u_{ij}(x,t)&=&
 \D \frac {\al _i-\al _j}{\gamma  _i -\gamma  _j}\langle g_x^{H^x}g_{t}^{H^t}\Phi_{i0},
Bg_x^{H^x}g_{t}^{H^t}\Psi_{j0}\rangle\vspace{1mm}  \\
&=&\D \frac {\al _i-\al _j}{\gamma  _i -\gamma  _j}\langle g_{t}^{H^t}g_x^{H^x}\Phi_{i0},
Bg_{t}^{H^t}g_x^{H^x}\Psi_{j0}\rangle,\ 1\le i\ne j\le 4,
\ea \]
where the initial values ${\Phi}_{i0}$ and ${\Psi}_{i0}$ 
of $\Phi_{i}$ and $\Psi_i$, $1\le i\le 4$,
can be taken to be any arbitrary constant vectors of $\R ^N$.

In a word, solutions of 
two binary constrained flows (\ref{eq:xcfof4NWIEs}) and (\ref{eq:tcfof4NWIEs}) 
can lead to involutive solutions of the 
multi-wave interaction equations (\ref{eq:4MWIEs}). 
More importantly, such involutive solutions
show us 
an explicit B\"acklund transformation from 
the multi-wave interaction equations (\ref{eq:4MWIEs})
to two finite-dimensional Liouville integrable Hamiltonian systems 
determined by (\ref{eq:defofH^x}) and (\ref{eq:defofH^t}) 
and thus  
the integrability by quadratures for the multi-wave interaction equations (\ref{eq:4MWIEs}).

\section{Summary and Conclusion}
\label{sec:summary_manila00}
\setcounter{equation}{0}

The multi-wave interaction equations 
(\ref{eq:4MWIEs}) have been taken as an illustrative example
of binary nonlinearization.
Binary symmetry constraints (\ref{eq:bscof4MWIEs}), 
containing arbitrary distinct constants $\gamma_1,\gamma_2,\gamma_3,\gamma_4$,
and arbitrary non-zero constants $\mu_1,\mu_2,\mu_3,\mu_4$,
were proposed for the multi-wave interaction equations (\ref{eq:4MWIEs}).
Two finite-dimensional Liouville integrable Hamiltonian systems 
determined by (\ref{eq:defofH^x}) and (\ref{eq:defofH^t}),
resulted from two binary constrained flows, determine a kind of involutive solutions to  
the multi-wave interaction equations (\ref{eq:4MWIEs}).
If we take $B=I_N$, and 
\[\Gamma =V_0, \ \textrm{i.e.},\  \textrm{diag}(\gamma_1,\gamma_2,\gamma_3,\gamma_4)=\textrm{diag}(\beta_1,\beta_2,\beta_3,
\beta _4),\] 
then the result established here will cover all results presented in Ref. \cite{ShiZ-JMP2000}.
This also implies that our result leads to a larger class of finite-dimensional 
integrable Hamiltonian systems.

The theory of binary nonlinearization 
shows that binary symmetry constraints in the continuous case decompose 
soliton equations (PDEs)
into binary constrained flows (ODEs).
The $r$-matrix formulation can be used to expose 
the Liouville integrability of binary constrained flows.
The resulting constraints on the potentials
also present a kind of involutive solutions to soliton equations
and thus eventually show the integrability by quadratures for soliton equations.
The whole theory can also be applied to many other types of soliton equations. 

We point out that in the case on $2\times 2$ traceless matrix spectral problems,
binary nonlinearization will be reduced to mono-nonlinearization, if we take
the reduction \[
(\psi_{1j},\psi_{2j})^T = (-\phi_{2j},\phi_{1j})^T,
\ 1\le j\le N.\]
That is to say, 
in such a case, all results achieved through mono-nonlinearization 
can be obtained from the results achieved through 
binary nonlinearization. Therefore, in this sense, binary nonlinearization is much 
broader and more systematic. 
 
We have seen that symmetry constraints yield nonlinear constraints on potentials
of soliton equations, and put linear spectral problems (linear with respect to 
eigenfunctions) into nonlinear binary constrained flows (nonlinear again with
respect to eigenfunctions), which seemly makes it more complicated 
to solve soliton equations. 
However, since spectral problems are overdetermined, 
to guarantee the existence of eigenfunctions of 
spectral problems, one needs compatability conditions 
which are often nonlinear partial differential equations. 
The symmetry property of the constraints brings us 
the Liouville integrability for nonlinear binary constrained flows. 
Therefore, the symmetry property makes up for the complexity of nonlinearization
while manipulating binary nonlinearization.

Of particular interest in the study of binary symmetry constraints are to create new classical integrable Hamiltonian systems 
which supplement the known classes of classical integrable systems
and to expose the integrability by 
quadratures for soliton equations by using 
binary constrained flows.
We mention that high-order symmetry constraints with involved Lie-B\"acklund 
symmetries having non-degenerate Hamiltonians can also be carried out for soliton equations without much difficulty, but the case of degenerate Hamiltonians 
needs some particular consideration \cite{LiM-CSF2000}. 
The rigorous considerations on binary symmetry constraints and two-binary symmetry constraints,
especially on variable symplectic structures of binary constrained flows, 
are being expected to be made in a future publication.  

\section*{Acknowledgments}
This work was supported by a grant from the Research Grants Council of 
Hong Kong 
Special Administrative Region, China (Project no. 9040466), and
a grant from the City University of Hong Kong (Project no. 7001041).

\end{document}